\pdfoutput=1
\documentclass[a4paper,11pt]{article}
\usepackage{jheppub} % for details on the use of the package, please
\usepackage[usenames,dvipsnames]{xcolor}
\usepackage{amsmath,amssymb,bbold,epsfig}
\usepackage{amsfonts}
\usepackage{slashed}
\usepackage{dsfont}
\usepackage{graphicx,color}
\usepackage{bbm}
\usepackage{bm}
\usepackage[normalem]{ulem}
\definecolor{lightgray}{gray}{0.85}

\vspace*{-1.5cm}
\title{Damping of neutrino oscillations, decoherence and 
the lengths of neutrino wave packets}
\author{Evgeny~Akhmedov and Alexei Y. Smirnov}
\affiliation{
Max-Planck-Institut f\"ur Kernphysik, Saupfercheckweg 1, \\69117 Heidelberg,
Germany
}
\emailAdd{akhmedov@mpi-hd.mpg.de}
\emailAdd{smirnov@mpi-hd.mpg.de}

\abstract{
Spatial separation of the wave packets (WPs) of neutrino mass eigenstates 
leads to decoherence and damping of neutrino oscillations. Damping can also 
be caused by finite energy resolution of neutrino detectors or, in the case 
of experiments with radioactive neutrino sources, by finite width of the 
emitted neutrino line. We study in detail these two types of damping effects 
using reactor neutrino experiments and experiments with radioactive $^{51}$Cr 
source as examples. We demonstrate that the effects of 
decoherence by WP separation can always be incorporated into a modification 
of the energy resolution function of the detector and so are intimately 
entangled with it.  We estimate for the first time the lengths $\sigma_x$ 
of WPs of reactor neutrinos and neutrinos from a radioactive $^{51}$Cr source. 
The obtained values, $\sigma_x = (2\times 10^{-5} - 1.4\times 10^{-4})$ cm,  
are at least six orders of magnitude larger than the currently available 
experimental lower bounds. We conclude that effects of decoherence by WP separation 
cannot be probed in reactor and radioactive source experiments. 
}

\begin{document}
\maketitle
\flushbottom

\newcommand{\be}{\begin{equation}}
\newcommand{\ee}{\end{equation}}
\newcommand{\bea}{\begin{eqnarray}}
\newcommand{\eea}{\end{eqnarray}}

%=======================================================================
\section{\label{sec:intro}Introduction}
%=======================================================================
Recently, there has been an increased interest in studying the manifestations 
of the wave packet (WP) nature of neutrinos in neutrino oscillations. 
In a number of papers the possibility of probing 
quantum decoherence effects due to the separation of the WPs of different 
neutrino mass eigenstates composing an emitted flavour-state neutrino  
was discussed. In particular, possible manifestations of neutrino WP 
separation in reactor experiments \cite{DayaBay:2016ouy,deGouvea:2020hfl,
deGouvea:2021uvg,JUNO:2021ydg,Arguelles:2022bvt} 
and experiments with radioactive neutrino sources 
\cite{Arguelles:2022bvt} have been investigated   
(for earlier and related discussions, including those for other neutrino 
experiments, see e.g.\ \cite{Ohlsson:2000mj,
Blennow:2005yk,Kayser:2010pr,Akhmedov:2012uu,Jones,Coloma:2018idr,
Coelho:2017zes,BalieiroGomes:2018gtd,
Naumov:2021vds}).  

In \cite{DayaBay:2016ouy} the Daya Bay collaboration has analyzed their 
reactor neutrino data taking into account possible decoherence effects due to 
the finite momentum spread $\sigma_p$ of the neutrino WPs and treating  
$\sigma_p/p$, along with $\sin^2 2\theta_{13}$ and $\Delta m_{32}^2$, as a 
free parameter. The relative momentum uncertainty was 
thus assumed to be momentum-independent. 
Their analysis produced the constraint 
$\sigma_p/p < 0.23$ at 95\% C.L. Note that 
for typical momenta of reactor neutrinos $p\simeq 3$\,MeV 
this approximately corresponds to the length of the neutrino WP 
$\sigma_x\simeq 1/\sigma_p\gtrsim  
2.8\times 10^{-11}$\,cm. 

The authors of \cite{deGouvea:2020hfl,deGouvea:2021uvg} have carried out a 
similar analysis of the data of the Daya Bay, RENO and in 
\cite{deGouvea:2021uvg} also of the KamLAND reactor neutrino experiments, 
but using the length of neutrino WP $\sigma_x$ rather than $\sigma_p/p$ 
as a fit parameter. From their combined fit they found the lower bound 
\be
\sigma_x>2.1\times 10^{-11}\,{\rm cm}~~(90\%\; {\rm C.L.}).
\label{eq:combined}
\ee
In \cite{deGouvea:2020hfl}  
also the sensitivity of the future medium-baseline JUNO experiment to 
decoherence effects due to 
neutrino WPs separation was considered. 
It was found that JUNO would be able to 
improve the bound of eq.~(\ref{eq:combined}) by an order of magnitude. 
The JUNO collaboration itself, following the analysis of \cite{Blennow:2005yk}, 
has studied the expected sensitivity of their experiment to various mechanisms 
of damping of neutrino oscillations, including damping due to 
neutrino WP separation \cite{JUNO:2021ydg}. They concluded 
that JUNO should be able to 
set the limits $\sigma_p/p<1.04\times 10^{-2}$ and 
$\sigma_x>2.3\times 10^{-10}$\,cm, both at 95\% C.L.  

In \cite{Arguelles:2022bvt} possible effects of decoherence 
by neutrino WP separation on the searches for oscillations of $\nu_e$ and 
$\bar{\nu}_e$ to sterile neutrinos $\nu_s$  
were considered. The authors discussed the tension 
between the results of short-baseline reactor experiments, which 
put constraints on $\nu_e\to \nu_s$ oscillations, and the BEST 
radioactive neutrino source experiment 
\cite{Barinov:2021asz,Barinov:2021mjj,Barinov:2022wfh}, which claimed a 
positive signal. Assuming that the actual value of $\sigma_x$ 
coincides with the lower bound $2.1\times 10^{-11}$\,cm given in 
eq.~(\ref{eq:combined}), the authors argued that, due to possible damping 
effects, the results of BEST and of the reactor experiments can be 
reconciled with each other. The value of $\sigma_x$ used in 
\cite{Arguelles:2022bvt} is, however, unrealistic. 
We shall discuss ref.~\cite{Arguelles:2022bvt} in more detail in 
section~\ref{sec:discuss}. 

How does the separation of neutrino WPs occur? 
Neutrinos of different mass $m_i$ propagate with different group 
velocities $v_{gi}=\partial E_i/\partial p_i$, and for ultra-relativistic or 
almost degenerate in mass neutrinos 
their difference satisfies 
\be
\frac{\Delta v_g}{v_g}\simeq \frac{\Delta m^2}{2E^2}\,.
\label{eq:Deltavg}
\ee 
Because of this velocity difference and of the finite lengths of the neutrino WPs, 
the overlap of the WPs of different neutrino mass eigenstates composing an 
emitted neutrino flavor state will decrease with time.  
If neutrinos propagate sufficiently long distance, 
these WPs will completely separate.  

Neutrino oscillations are a quantum mechanical (QM) interference phenomenon; 
separation of the WPs of different neutrino mass eigenstates will suppress  
their interference in the oscillation probability and therefore will 
damp the oscillations. Requiring the spatial separation of the WPs to be
 smaller than the length of their WPs, 
one finds the constraint on the distance $L$ traveled by the neutrinos: 
\be
L\,<\,L_{\rm coh}\equiv \frac{v_g}{\Delta v_g}\sigma_x\,,
\label{eq:cond}
\ee 
where $\sigma_x\simeq v_g/\sigma_E$ is the length of the neutrino WP. 
Here $\sigma_E$ is  
the intrinsic QM uncertainty of the neutrino energy related to the 
localization of its production and detection processes.
Taking into account eq.~(\ref{eq:Deltavg}), for ultra-relativistic neutrinos 
condition (\ref{eq:cond}) yields 
\be 
\frac{\sigma_E}{E}< \frac{1}{2\pi}\frac{l_{\rm osc}}{L}\,, 
\label{eq:propCoh1}
\ee
where $l_{\rm osc}\equiv\frac{4\pi E}{\Delta m^2}$ is the neutrino oscillation 
length.

Quantum decoherence due to WP separation is not the only possible reason for 
damping of neutrino oscillations. The damping may also occur due to the averaging 
over the baseline $L$ related to the uncertainties of the coordinates of the 
neutrino emission and absorption points, both due to the finite spatial extensions 
of the elementary production and detection processes and, more importantly, 
due to the 
macroscopic sizes of the neutrino source 
and detector. For neutrino oscillations to be observable, the corresponding 
averaging regions should be small compared to the neutrino oscillation length, 
i.e.\ neutrino production and detection should be sufficiently well localized.

There is yet another possible source of damping of neutrino oscillations: 
finite energy resolution of the detector $\delta_E$. 
The experiment will only be able to see the oscillations if the oscillation 
phase 
$\phi(E)=\frac{\Delta m^2}{2E}L$ varies little over the energy interval 
$\delta_E$. 
Requiring that $|\phi(E)-\phi(E+\delta_E)|<1$, we find that 
$\delta_E$ must satisfy the inequality   
\be 
\frac{\delta_E}{E}< \frac{1}{2\pi}\frac{l_{osc}}{L}\,.
\label{eq:enAver}
\ee
Note that this condition coincides with that in eq.~(\ref{eq:propCoh1}) 
with $\sigma_E$ replaced by $\delta_E$. Additional damping effects can be 
related to energy binning of the data.  

Clearly, averaging over $L$ and $E$, inherent in any neutrino oscillation 
experiment, may damp neutrino oscillations and thus mimic quantum 
decoherence by WP separation. 
It is therefore mandatory to carefully examine these averaging effects when 
trying to probe experimentally decoherence due to the WP nature of neutrinos.  
In the present paper we consider in detail the damping of neutrino 
oscillations due to decoherence by WP separation and due to 
energy averaging related to finite experimental energy resolution. We 
study the entanglement of 
these two effects 
for reactor neutrino experiments and experiments 
with radioactive neutrino sources, 
though our results have broader applicability. 
We also estimate the spatial lengths of neutrino WPs in reactor and source 
experiments and the corresponding intrinsic QM neutrino energy uncertainties.
To the best of our knowledge, these are the first 
consistent estimates of these quantities. 

%=============================================================================
\section{\label{sec:WP}Separation of wave packets vs. energy averaging}
%=============================================================================
%=============================================================================
\subsection{\label{sec:WP1}
Wave packet approach and the oscillation damping factor}
%=============================================================================

In the WP approach, the probability of $\nu_\alpha\to \nu_\beta$ 
oscillations in vacuum can be written in the following form 
(see e.g.\ \cite{Akhmedov:2009rb}): 
\be 
P_{\alpha\beta}(\bar{E},L) = \sum_{i,k} U^*_{\alpha i} U_{\beta i} 
U_{\alpha k} U_{\beta k}^*\, 
I_{ik}(\bar{E},L)\,. 
\label{eq:P1} \ee 
Here $L$ is the baseline and $U$ is the leptonic mixing matrix. 
The quantity $I_{ik}(\bar{E},L)$ can be written as \cite{Akhmedov:2009rb} 
\be 
I_{ik}(\bar{E},L)= 
\int dE\, |f(E,\bar{E})|^2 e^{-i\frac{\Delta m_{ik}^2}{2E}L}\,. 
\label{eq:Iik1} 
\ee 
Here 
\be 
f(E,\bar{E})\equiv 
f_{S}(E)f_{D}^*(E)\,, 
\label{eq:f1} 
\ee 
where $f_{S}(E)$ and $f_{D}(E)$ 
are the WPs of the produced and detected neutrinos in energy representation 
(i.e.\ the amplitudes of the corresponding energy distributions) and 
$\bar{E}$ is the mean energy of the neutrino WP. The quantity $f(E,\bar{E})$ 
is normalized according to 
\be 
\int dE\, |f(E,\bar{E})|^2=1\,, 
\label{eq:norm2} 
\ee which gives 
$I_{ii}(\bar{E},L)=1$. The mean energy is defined as $\bar{E}=\int dE\, E 
|f(E,\bar{E})|^2$. The function $f(E,\bar{E})$ is an 
effective neutrino WP in energy space, which takes the quantum nature of both 
neutrino production and detection into account \cite{Akhmedov:2009rb}. 
The energy distribution amplitudes 
$f_{S}(E)$ and $f_{D}(E)$ 
have peaks of widths $\sigma_{E}^S$ and $\sigma_{E}^D$, 
respectively. As $f(E,\bar{E})$ is the product of 
$f_S(E)$ and $f_D(E)$,
it has a peak of width $\sigma_E\simeq 
\min\{\sigma_{E}^S,\,\sigma_{E}^D\}$ with maximum at or near $\bar{E}$.  
The quantity $I_{ik}(\bar{E},L)$ 
depends on the oscillation phase
and on the degree of overlap of the WPs of different neutrino mass 
eigenstates in momentum and coordinate spaces and thus encodes possible 
WP-related quantum decoherence effects. 

In the present paper we assume that the microscopic localization conditions 
for neutrino production and detection are fulfilled, 
i.e.\ the corresponding averaging over $L$ can be neglected. 
For elementary neutrino production and detection processes these localization 
conditions require \cite{Kayser:1981ye, Akhmedov:2009rb,Giunti:1997wq, 
Giunti:2002xg,Beuthe:2001rc} 
\be
\frac{\Delta m_{ik}^2}{2E}\ll \sigma_E\,.
\label{eq:cond1}
\ee
When deriving eqs.~(\ref{eq:Iik1}) and~(\ref{eq:f1})  
from the general expressions given in \cite{Akhmedov:2009rb}, 
we assumed this condition to be satisfied.% 
\footnote{
If this condition were not met, the arguments of $f_S(E,\bar{E})$ 
and $f_D(E,\bar{E})$ in eq.~(\ref{eq:f1}) would differ by 
$\Delta m_{ik}^2/2E$, and the oscillations would be damped.  
}  
We also found it convenient to go from the usual integration over momentum to 
integration over energy. It should be stressed, however, that the same result 
could be obtained using coordinate-space neutrino WPs \cite{Akhmedov:2009rb}.  

It is convenient to rewrite the expression for $I_{ik}(\bar{E},L)$ by   
pulling the phase factor $\exp(-i\frac{\Delta m_{ik}^2}{2E^{}}L)$
taken at $E=\bar{E}$ out of the integral in eq.~(\ref{eq:Iik1}).
This yields 
\be I_{ik}(\bar{E},L)= 
\exp\Big(\!-i\frac{\Delta m_{ik}^2}{2\bar{E}}L\Big)\,
D_{ik}(\bar{E},L) 
\label{eq:Iik3} 
\ee 
with 
\be
D_{ik}(\bar{E},L)\simeq \int dE 
|f(E,\bar{E})|^2 e^{i\frac{\Delta m_{ik}^2}{2\bar{E}^2}(E-\bar{E})L}\,.  
\label{eq:Iik4}
\ee 
{}From the fact that $|f(E,\bar{E})|^2$ has a peak at or near $\bar{E}$ of width 
$\sigma_E$ 
it follows that the main contribution to the integral 
defining 
$D_{ik}(\bar{E},L)$ 
comes from the region 
$|E-\bar{E}|\lesssim \sigma_E$. 
It is then easy to see that for 
\be 
L\ll L_{{\rm coh},ik}\equiv
\frac{2\bar{E}^2}{\Delta m_{ik}^2}\sigma_E^{-1}\simeq 
\frac{2\bar{E}^2}{\Delta m_{ik}^2}\sigma_x\,
\label{eq:Lcoh}
\ee
(which corresponds to the absence of WP separation) one has 
$D_{ik}(\bar{E},L)=1$. Eqs.~(\ref{eq:Iik3}) and~(\ref{eq:P1})
then yield the standard master formula for neutrino oscillations 
in vacuum: $P_{\alpha\beta}(\bar{E},L)=P_{\alpha\beta}^0(\bar{E},L)$, where 
\be
P_{\alpha\beta}^0(E,L)
\equiv\sum_{i,k} U^*_{\alpha i} U_{\beta i} U_{\alpha k} 
U_{\beta k}^*\,e^{-i\frac{\Delta m_{ik}^2}{2E}L}\,. 
\label{eq:P0}
\ee
In the opposite case, $L\gg L_{{\rm coh},ik}$ (complete decoherence by WP 
separation), the integrand of eq.~(\ref{eq:Iik3}) contains a fast oscillating 
phase factor which suppresses  the quantities $D_{ik}(\bar{E},L)$ 
with $i\ne k$. This gives $D_{ik}(\bar{E},L)=\delta_{ik}$, 
and eq.~(\ref{eq:P1}) yields 
\be
P_{\alpha\beta}(\bar{E},L) = \sum_{i} |U_{\alpha i}|^2 |U_{\beta i}|^2\,,
\label{eq:P2}
\ee
i.e.\ the oscillations are fully damped. The quantity $D_{ik}(\bar{E},L)$ 
with $i\ne k$ is thus the oscillation damping factor. 
Note that the 
fully decoherent result (\ref{eq:P2}) would also follow from the standard 
oscillation probability (\ref{eq:P0}) upon averaging of all the oscillatory 
terms. 

For illustration, we consider the energy-space neutrino WP of Gaussian form: 
\be
|f(E,\bar{E})|^2=
\frac{1}{\sqrt{2\pi}\sigma_E}e^{-\frac{(\bar{E}-E)^2}{2\sigma_E^2}}\,. 
\label{eq:Gauss1}
\ee
Substituting this into eq.~(\ref{eq:Iik4}), from eqs. (\ref{eq:P1}) and 
(\ref{eq:Iik3}) we find 
\be
P_{\alpha\beta}(\bar{E},L) = \sum_{i,k} U^*_{\alpha i} U_{\beta i} U_{\alpha k} 
U_{\beta k}^*\,
\exp\Big(-i\frac{\Delta m_{ik}^2}{2\bar{E}}L\Big)\, 
D_{ik}(\bar{E},L)\,,
\label{eq:P1a}
\ee
where 
\be
D_{ik}(\bar{E},L)=e^{-\frac{1}{2}\left(\frac{L}{L_{{\rm coh}, ik}}
\right)^2} 
\label{eq:Damp}
\ee
is the Gaussian the damping factor 
often used in the literature on neutrino oscillations.

%=============================================================================
\subsection{\label{sec:WP2}Finite energy resolution and energy averaging 
in reactor experiments}
%=============================================================================
The number of neutrino events per unit time in a neutrino experiment   
can be written as 
\be
N(E_r)={\cal N} 
\int d\bar{E} \phi_\alpha(\bar{E})P_{\alpha\beta}(\bar{E}, L)
\sigma_\beta(\bar{E}) R(E_r,\bar{E})\,, 
\label{eq:numbev1}
\ee
where ${\cal N}$ is the number of the target particles in the detector, 
$\bar{E}$ is the discussed above mean energy of the neutrino WPs, 
$E_r$ is the reconstructed neutrino energy, $\phi_\alpha(\bar{E})$ is the 
flux of the neutrinos, 
initially produced as $\nu_\alpha$, impinging on the detector, 
$\sigma_\beta(\bar{E})$ is the cross section of detection of 
$\nu_\beta$ and $R(E_r,\bar{E})$ is the energy resolution function of the 
detector.  The oscillation probability $P_{\alpha\beta}(\bar{E},L)$ contains 
a damping factor which takes into account possible effect of decoherence by WP 
separation. For reactor experiments, $\alpha=\beta=e$, but we want to keep 
our discussion more general at this point, so that it also apply to other 
beam experiments. 

The oscillation probability (\ref{eq:P1}) with $I_{ik}(\bar{E},L)$ from 
eq.~(\ref{eq:Iik1}) can be written as 
\be
P_{\alpha\beta}(\bar{E},L) = \int dE |f(E,\bar{E})|^2 P_{\alpha\beta}^0(E,L)\,,
\label{eq:P2a}
\ee
where $P_{\alpha\beta}^0(E,L)$ is 
given in eq.~(\ref{eq:P0}). Substituting 
(\ref{eq:P2a}) into (\ref{eq:numbev1}) and changing the order of integrations, 
we obtain  
\be
N(E_r)={\cal N} \int dE\, P_{\alpha\beta}^0(E,L)
\int d\bar{E}\, |f(E,\bar{E})|^2 
\phi_\alpha(\bar{E})\sigma_\beta(\bar{E})R(E_r,\bar{E})
\,. 
\label{eq:numbev3}
\ee
Next, we notice that,  
while $|f(E,\bar{E})|^2$ as a function of $\bar{E}$ has a peak of 
small width $\sigma_E$ at $\bar{E}=E$ or very close to this value, 
the flux $\phi_\alpha(\bar{E})$ and the cross section $\sigma_\beta(\bar{E})$ 
are smooth functions that change very little over the intervals $\Delta \bar{E}
\sim \sigma_E$;  therefore to a very good accuracy they can be replaced by 
their values at $\bar{E}=E$ and pulled out of the inner integral.  
This yields 
\be
N(E_r)={\cal N} \int dE\, \phi_\alpha(E)P_{\alpha\beta}^0(E,L)
\sigma_\beta(E)
\tilde{R}(E_r,E)\,, 
\label{eq:numbev4}
\ee
where
\be
\tilde{R}(E_r,E) = 
\int d\bar{E}\, R(E_r,\bar{E})|f(E,\bar{E})|^2 \,. 
\label{eq:tildeR}
\ee
Comparing this result with (\ref{eq:numbev1}), we see that they differ in two 
respects: first, the integrand of (\ref{eq:numbev4}) contains the standard 
oscillation probability $P^0_{\alpha\beta}(E,L)$ (which is free from any 
decoherence effects) rather than the full probability $P_{\alpha\beta}(E,L)$ 
of eq.~(\ref{eq:P1}); second, (\ref{eq:numbev4}) contains an effective 
energy resolution function $\tilde{R}(E_r,E)$ rather than the true one. This 
shows that effects of quantum decoherence due to WP separation can be 
incorporated into a modification of the energy resolution function of 
the detector and so are intimately entangled with it. 

How much is the detector resolution modified by including the possible quantum 
decoherence effects into it? 
To illustrate this, we consider the case where 
both the energy-space neutrino WP and the resolution function $R(E_r,E)$
are of Gaussian form, that is, $|f(E,\bar{E})|^2$ is given by 
eq.~(\ref{eq:Gauss1}) and\,% 
\footnote{Gaussian energy resolution functions with 
energy-dependent widths $\delta_E(E)$ are often used by experimentalists. 
Here for simplicity we take $\delta_E$ to be energy 
independent. }
\be
R(E_r,\bar{E})=
\frac{1}{\sqrt{2\pi}\delta_E}e^{-\frac{(E_r-\bar{E})^2}{2\delta_E^2}}\,.
\label{eq:Gauss2}
\ee
Substituting 
(\ref{eq:Gauss1}) and (\ref{eq:Gauss2}) into eq.~(\ref{eq:tildeR}) and 
extending the energy integration to the interval $(-\infty,\infty)$ (which is 
justified because the integrand has peaks of small width at positive values 
of $\bar{E}$), we obtain 
\be
\tilde{R}(E_r, E)=
\frac{1}{\sqrt{2\pi(\delta_E^2+\sigma_E^2)}}
e^{-\frac{(E_r-E)^2}{2(\delta_E^2+\sigma_E^2)}}\,.
\label{eq:Gauss3}
\ee
For $\delta_E\gg\sigma_E$, the effective energy resolution function 
$\tilde{R}(E_r,E)$ essentially coincides with the true one. This means 
that in this case quantum decoherence by WP separation can be 
completely neglected, and whether or not the oscillations will be damped 
will be determined by condition (\ref{eq:enAver}). 
The separation of WPs may only be probed by the experiment if 
$\sigma_E\gtrsim \delta_E$. 

%========================================================================== 
\subsection{\label{sec:source1}Radioactive source experiments} 
%========================================================================== 

In experiments of this type neutrinos are emitted by a $\beta$-radioactive 
source placed inside 
or near the detector. Usually, $\beta$-decay by electron capture is considered 
for the source; in this 
case neutrinos have (quasi)monoenergetic spectra.
By now, source experiments have been performed by the 
GALLEX \cite{GALLEX:1997lja,Kaether:2010ag}, 
SAGE \cite{SAGE:1998fvr,Abdurashitov:2005tb}
and more recently BEST 
\cite{Barinov:2021asz,Barinov:2021mjj,Barinov:2022wfh} 
collaborations. In all these cases $^{71}$Ga was used as the target, and 
$^{71}$Ge atoms produced as a result of neutrino capture by gallium were 
counted. Neutrino energy was not measured, and therefore, unlike in reactor 
experiments, the detection rates 
had no dependence on 
the neutrino energy resolution.%
\footnote{There have been suggestions to use neutrino-electron scattering 
rather than neutrino capture by atomic nuclei as a detection process in source 
experiments, see e.g.\ \cite{Vergados:2011na,Borexino:2013xxa}. In those cases 
the experimental energy resolution functions would have to be taken into 
account. With minor modifications, the formalism developed in 
section~\ref{sec:WP2} would then apply.}
GALLEX and SAGE measured the coordinate-averaged neutrino flux, whereas 
BEST comprises an inner and an outer targets and therefore has some 
coordinate sensitivity. 

Consider a neutrino source experiment with $e$-capture radioactive nuclei as 
a source and counting of daughter nuclei as the means of neutrino detection. 
Experiments of this 
type look for $\nu_e$ disappearance and therefore the observed signal depends 
on $P_{ee}(E,L)$.  We shall be assuming that the source is small compared with 
both the detector size and the neutrino oscillation length; the source can 
then be considered as pointlike. We shall put the origin of coordinates at the 
source. The detector may consist of one or more target volumes 
$V_i$. The number of events per unit time in the $i$th volume can then be 
written as 
\be
N_i(t)=
n_0\int_{V_i}
d^3r \int\,d\bar{E}\,\phi_e(\bar{E},r;t)\sigma_e(\bar{E})P_{ee}(\bar{E},r)\,,
\label{eq:Nev1}
\ee
where $n_0$ is the number density of the target nuclei and  
\be
\phi_e(\bar{E},r;t)=\frac{\Phi_e(\bar{E},t)}{4\pi r^2}\,
\label{eq:phi1}
\ee
is the flux of $\nu_e$ at the distance $r$ from the source at the time $t$. 
The quantity $\Phi_e(\bar{E},t)$ can be written as 
\be
\Phi_e(\bar{E},t)=\Gamma_0 N(t)
S(\bar{E})\,, 
\label{eq:phi2}
\ee
where $\Gamma_0$ is the electron capture rate of the source atoms, 
$N(t)$ is the number of the source atoms at the time $t$ (which decreases with 
time following the exponential decay law) and 
$S(\bar{E})$ is the 
normalized spectrum of the produced neutrinos. It 
is characterized by a width $\delta_{El}$ which usually exceeds significantly 
the natural linewidth and is determined 
by a number of inhomogeneous broadening effects, such as Doppler broadening.  
Note that inhomogeneous broadening leads to the spread of the mean energies 
$\bar{E}$ of the WPs of the emitted neutrinos due to 
the individual emitters being in slightly different conditions. 
We shall discuss inhomogeneous broadening (as well as homogeneous broadening 
which affects $f(E,\bar{E})$) 
in more detail in section~\ref{sec:source2} below. 

We shall now follow essentially the same steps as in section~\ref{sec:WP2}. 
Using eqs.~(\ref{eq:phi1}) and (\ref{eq:phi2}) in 
(\ref{eq:Nev1}), we obtain
\be
N_i(t)=n_0\Gamma_0 N(t)
\int_{V_i} \frac{d^3r}{4\pi r^2}\,{\cal F}(r)\,,
\label{eq:Nev2}
\ee
where 
\be
{\cal F}(r)=\int\,d\bar{E}\,S(\bar{E})\sigma_e(\bar{E})\,P_{ee}(\bar{E},r)\,.
\label{eq:calF}
\ee
Let us consider the quantity ${\cal F}(r)$. 
Substituting in (\ref{eq:calF})  the expression for $P_{ee}(\bar{E},r)$ from 
(\ref{eq:P2a}) and changing the order of integrations over $E$ and 
$\bar{E}$ yields 
\be
{\cal F}(r)=
\int\,dE\,P^0_{ee}(E,r) 
\int\,d\bar{E}\,
S(\bar{E})
\sigma_e(\bar{E})
\,|f(E,\bar{E})|^2\,.
\label{eq:Nev3}
\ee
The cross section $\sigma_e(\bar{E})$ changes very little over the energy 
intervals 
$\Delta\bar{E}\sim \sigma_E$, and therefore it can be 
replaced by its value at $\bar{E}=E$ and pulled out of the inner integral in 
(\ref{eq:Nev3}). This gives 
\be
{\cal F}(r)=\int\,dE\,P^0_{ee}(E,r)\sigma_e(E)
\tilde{S}(E)\,,
\label{eq:Nev4}
\ee
where
\be
\tilde{S}(E)
=\int\,d\bar{E}\,
S(\bar{E})|f(E,\bar{E})|^2\,.
\label{eq:tildePhi}
\ee
All the effects of decoherence by WP separation are now incorporated into a 
modification of the neutrino spectrum 
$S(E)$, 
which is replaced by the effective spectrum $\tilde{S}(E)$. 
Note that for source experiments the effective spectrum 
$\tilde{S}(E)$ plays essentially the same role as the 
effective energy resolution $\tilde{R}(E_r,E)$ plays for reactor experiments 
(cf.~eq.~(\ref{eq:tildeR})). 

Obviously, if the energy width $\sigma_E$ of the neutrino WP 
satisfies $\sigma_E\ll \delta_{El}$, one can replace 
$S(\bar{E})$ by 
$S(E)$ 
in eq.~(\ref{eq:tildePhi}) and pull it out of the integral, 
which gives $\tilde{S}(\bar{E})=S(E)$.  
Eqs.~(\ref{eq:Nev2}) and (\ref{eq:Nev4}) then yield the usual expression for 
the event rate in the absence of WP decoherence. 
The damping will then only depend on the width of the neutrino spectrum 
$\delta_{El}$. 

In the opposite limit, $\sigma_E\gg \delta_{El}$, one can instead 
pull out of the integral the factor $|f(E,\bar{E})|^2$ 
at $\bar{E}=E_0$, where $E_0$ is the central energy of the neutrino spectrum. 
Eq.~(\ref{eq:tildePhi}) then  
gives 
$\tilde{S}(E)=|f(E,E_0)|^2$. 
This yields  
\be
{\cal F}(r)=\int\,dE\,P^0_{ee}(E,r)\sigma_e(E)|f(E,E_0)|^2
\simeq \sigma_e(E_0)P_{ee}(E_0,r)\,,
\label{eq:Int1}
\ee
where we have used eq.~(\ref{eq:P2a}). The oscillation probability 
$P_{ee}(E_0,r)$ here fully takes into account possible WP decoherence 
effects. 

As before, we illustrate the above points by using the Gaussian form of the 
neutrino WP (\ref{eq:Gauss1}) and assuming that the spectrum of the neutrinos 
produced by the radioactive source is also Gaussian: 
\be
S(\bar{E})
=\frac{1}{\sqrt{2\pi}\delta_{El}}
e^{-\frac{(\bar{E}-E_0)^2}{2\delta_{El}^2}}\,.
\label{eq:Gauss4}
\ee
Eq.~(\ref{eq:tildePhi}) then gives for the effective neutrino spectrum 
$\tilde{S}(E)$ 
the expression that coincides with the right-hand side of 
eq.~(\ref{eq:Gauss3}) with the replacements $\delta_E\to\delta_{El}$, 
$E_r\to E_0$.

%========================================================================== 
\section{\label{sec:lengths}Lengths of neutrino wave packets} 
%========================================================================== 

Let us now estimate the lengths of the neutrino WPs and the corresponding 
neutrino energy uncertainties $\sigma_E$ for reactor and radioactive source 
experiments.

The length of the WP of a neutrino produced in a decay or collision process 
is given by \cite{Giunti:2002xg,Beuthe:2001rc} 
\be
\sigma_x\simeq \frac{v_g-v_P}{\sigma_E}\,,
\label{eq:length1}
\ee
where $v_g$ is the velocity of the emitted neutrino and $v_P$ is the 
velocity of the neutrino source. For reactor and radioactive source 
experiments, the neutrino source is a highly non-relativistic nucleus and 
its velocity $v_P$ can be neglected compared with $v_g\simeq 1$ in 
eq.~(\ref{eq:length1}). The QM uncertainty of neutrino energy $\sigma_E$ 
is given by the inverse of the temporal duration of the production 
process $\sigma_t$: $\sigma_E\simeq \sigma_E^S=\sigma_t^{-1}$.%
\footnote{\label{ftnt:det} As was discussed in section~\ref{sec:WP1}, the quantity 
$\sigma_E$ of interest to us is actually the smaller between $\sigma_E^S$ 
and $\sigma_E^D$ which correspond, respectively, to neutrino production and 
detection. It can be shown that in the cases we consider this is actually 
$\sigma_E^S$, which we hereafter will simply denote $\sigma_E$.  
}

We shall consider two cases: 
\begin{enumerate}
\item[(i)] Particles accompanying neutrino production 
are not detected and do not interact with the surrounding medium. 

\item[(ii)] Some or all of the particles accompanying neutrino production 
are ``measured'', i.e.\ they are either directly detected or interact with  
the particles of the medium. 

\end{enumerate}

%=====================================================================
\subsection{\label{sec:reactor}Wave packets of 
reactor (anti)neutrinos}
%=====================================================================
	
Consider first neutrino emission in $\beta$-decays of nuclear fragments 
produced in fission reactions in nuclear reactors, 
\be
N\to N'+e^-+\bar{\nu}_e\,,
\label{eq:beta1}
\ee
where $N$ and $N'$ are the parent and daughter nuclei.

%=============================================================================
\subsubsection{\label{sec:parent}Neutrino WPs in the case of delocalized 
accompanying particles}
%=============================================================================

We start with the case when all the particles produced together with neutrino 
escape freely. In this case they do not affect the neutrino production time 
and the neutrino simply inherits the energy uncertainty of the parent unstable
 nucleus. If this nucleus is free or quasi-free, i.e.\ its interactions with 
the medium can be neglected, neutrino emission proceeds uninterrupted,  
and the characteristic emission time satisfies $\sigma_t\simeq \tau_N$, where 
$\tau_N$ is the mean lifetime of $N$. Thus, in this case 
$\sigma_E\simeq 1/\tau_N=\Gamma_N$, where $\Gamma_N$ is the decay width 
of the parent nucleus. 

If, however, $N$ experiences collisions with the particles 
of the medium and the average time interval between two successive  
collisions is shorter than the lifetime $\tau_N$, such interactions will 
lead to interruptions of coherent neutrino emission and therefore will 
increase its energy uncertainty $\sigma_E$. This effect is analogous to 
collisional broadening of photon emission lines in atomic physics. As we shall 
see, for neutrinos produced in $\beta$-decays of fission products in nuclear 
reactors collisional broadening is an important effect that has to be taken 
into account.  

We shall bear in mind that the duration of the processes of collision 
of the parent nucleus $N$ with the surrounding particles of the medium 
are very short compared to its lifetime $\tau_N$ and shall also assume that 
these collisions introduce uncontrollable random phases into the wave 
function of the emitted neutrino state. In this 
approach (known in atomic physics as the Lorentz--Van Vleck--Weisskopf 
approach, see e.g. ref.~\cite{Lorentz}), the energy 
width of the WPs of the emitted neutrino state obeys 
\be
\sigma_E\simeq \sqrt{\Gamma_N^2+1/t_N^2}\,,
\label{eq:broad}
\ee
where $t_N$ is the mean time interval between two successive collisions 
(mean free time). 

Let us estimate $t_N$. Immediately after the fission the kinetic energies of 
the produced nuclear fragments can be as large as about 100\,MeV.  
The fragments quickly thermalize on the time scale that is 
many orders of magnitude shorter than their lifetimes $\tau_N$ 
with respect to $\beta$-decay.%
\footnote{For short-lived fission fragments, the lifetimes 
are typically in the range of 
minutes to days.}
Therefore, at the moment of the decay the parent nuclei are in thermal 
equilibrium with the medium, and their velocities are determined by the 
temperature of the medium $T$. For a fragment of mass $m_N$ the average 
velocity is
\be
v_N = \sqrt{\frac{3T}{m_N}}.    
\label{eq:velN}
\ee
Taking for estimate  
$T \simeq 0.1$ eV (1160\,K) and $m_N \simeq 100$ GeV, we find 
\be
v_N \,\simeq\, 1.7 \times 10^{-6}c \,\simeq\, 5.2 \times 10^{4}\,{\rm cm/s}\,.
\label{eq:velN2}
\ee 

As the velocities of the fragments are small compared with the velocities of 
atomic electrons, they drag along the electrons of the parent atom of the 
fissile material and emerge as neutral or weakly ionized atoms.  
Their interactions with the medium are therefore  
described by atomic scattering cross sections $\sigma_{AA}$. 

Nuclear fuel of commercial reactors is usually composed of a mixture of 
$^{235}$U, $^{238}$U  $^{239}$Pu and $^{241}$Pu. For definiteness, 
we shall consider the scattering of nuclear fragments in pure $^{235}$U 
(this simplification will not affect our estimates significantly). 
The average time between two successive collisions is then   
\be
t_N \simeq \frac{1}{\sigma_{AA} n_U v_N}\,,
\label{eq:timeN}
\ee
where $n_U=4.9\times 10^{22}$ cm$^{-3}$ is the number density of uranium 
atoms. The mean free path of $N$, which actually 
determines its localization in the medium, is 
\be
X_N=v_N t_N=\frac{1}{\sigma_{AA} n_U}\,.
\label{eq:XN}
\ee

We take the radius of uranium atoms to be equal to their van der Waals 
radius $r_{\rm vdW} = 1.86 \times 10^{-8}$\,cm. 
For a crude estimate of $\sigma_{AA}$, we approximate it 
by the geometrical cross section: $\sigma_{AA} \simeq \pi (2r_{\rm vdW})^2$. 
This is expected to be a reasonable approximation because \mbox{(i) the} 
nuclear momenta satisfy 
$p_N r_{\rm vdW}\gg 1$, so that the WKB approximation applies, and (ii) when 
the atoms approach each other to distances $\lesssim 2r_{\rm vdW}$ they 
experience strong repulsion with a potential $U$ which exceeds significantly 
both the kinetic energy of $N$ ($\sim 0.15$\,eV) and the quantity 
$1/(m_N r_{\rm vdW}^2)\simeq 0.1$\,eV, and therefore the approximation of 
scattering on a rigid sphere should apply. Eqs.~(\ref{eq:timeN}) and 
(\ref{eq:velN2}) then give 
\be
t_N\simeq 
9\times 10^{-14}\,{\rm s}\,, 
\label{eq:timeN3}
\ee 
and $X_N\simeq 
5\times 10^{-9}$\,cm. As $t_N$ is much shorter than the lifetime of the 
decaying nucleus, eq.~(\ref{eq:broad}) yields 
\be
\sigma_E\simeq 
t_N^{-1}\simeq 
7.2\times 10^{-3}\,{\rm eV}\,.
\label{eq:sigmaElength2}
\ee
{}From (\ref{eq:length1}) we then find that in the limit when the particles 
accompanying neutrino production are completely delocalized  
\be
\sigma_x\simeq \frac{v_g}{\sigma_E}\simeq 
2.8\times 10^{-3}\,{\rm cm}\,.
\label{eq:sigma1}
\ee
Note that from eqs.~(\ref{eq:sigma1}),~(\ref{eq:sigmaElength2}) 
and~(\ref{eq:XN})
it follows that the expression for $\sigma_x$ can also be written as 
\be
\sigma_x\simeq X_N\frac{v_g}{v_N}\,.
\label{eq:length3}
\ee
That is, the length of the neutrino WP is given in this case by the mean free 
path of the parent nucleus magnified by a very large factor 
$v_g/v_N\simeq 6\times 10^{5}$.

%=============================================================================
\subsubsection{\label{sec:daughter}Effects of localization of the decay 
products}
%=============================================================================

Let us now take into account the interactions of the decay products 
accompanying neutrino emission with medium. We shall first consider the 
interactions of the daughter nucleus $N'$. 

Collisions of $N'$ with the atoms of the medium localize it within the 
spatial region of the size $X_{N'}$, which can be estimated similarly to the 
localization of $N$ (see eq.~(\ref{eq:XN})), using the same geometrical 
atom-atom scattering cross section 
$\sigma_{AA}$.%
\footnote{We assume here that the van der Waals radii of the 
fission fragments are of the same order of magnitude as that of uranium.} 
Thus, the mean free paths of the parent and daughter nuclei are of the same 
order of magnitude:
\be
X_{N'}\simeq X_N\simeq 
5\times 10^{-9}\,{\rm cm}\,.
\label{eq:Xprime}
\ee
The collisions of $N'$ with the surrounding atoms interrupt the process of its 
coherent emission. 
As $N'$ is produced in the same $\beta$-decay process in which the  neutrino 
is emitted, this will also affect the coherence time of neutrino production 
$\sigma_t$. 

Let us estimate the velocity $v_{N'}$ with which the daughter 
nucleus $N'$ is emitted. On average, the momenta 
of all the particles in the final state of the $\beta$-decay process 
(\ref{eq:beta1}) are roughly of the same order: $p_e\sim p_\nu\sim p_{N'}$.
For the typical reactor neutrino energy $E\simeq 3$\,MeV and $m_{N'}\sim 
100$\,GeV we find 
\be
v_{N'}= p_{N'}/m_{N'}\sim 3\times 10^{-5}c 
\simeq 10^6\,{\rm cm/s}\,.
\ee
For the mean time between two successive collisions of $N'$ we therefore obtain  
\be
t_{N'}=\frac{X_{N'}}{v_{N'}}\simeq 
5\times 10^{-15}\,{\rm s}\,.
\label{eq:timeNprime}
\ee
Comparing this with (\ref{eq:timeN3}), we find that $t_{N'}$ is about a factor 
of 20 smaller than $t_N$. 

Let us now consider effects of interaction with the medium of the electron 
produced in 
decay (\ref{eq:beta1}). 
The main effect of 
scattering of $\beta$-electrons on atoms is the ionization of the latter. 
For electron kinetic energies $E_e$ up to a few keV, the cross section of 
electron-impact ionization of uranium $\sigma_{eU}$ can be found in  
\cite{goswami}. Extrapolating it to MeV-scale energies basing on the results 
of ref.~\cite{gumus}), for electron energies $E_e\sim 3$\,MeV we 
find $\sigma_{eU}\simeq 1\times 10^{-18}\,{\rm cm}^2$.  
Therefore, for the mean free path of electrons we obtain  
$X_e=1/(\sigma_{eU}n_U)\simeq 2\times 10^{-5}$\,cm, which is a factor of 
$4\times 10^3$ larger than $X_N$ and $X_{N'}$. On the other hand, for 
typical electron momenta $p_e\sim 3$\,MeV their velocities are close to 1, 
and therefore the average time between two successive collisions of a 
$\beta$-decay electron is 
\be
t_e=X_e/v_e\simeq 7\times 10^{-16}\,{\rm s}\,.
\label{eq:timeE}
\ee

As the collisions of the parent nucleus $N$ and of the decay products in 
reaction (\ref{eq:beta1}) with the surrounding atoms of the medium lead to 
interruptions of their coherent propagation or emission, the time of the 
coherent production of neutrino is determined by the shortest among the 
coherence times $t_N$, $t_{N'}$ and $t_e$ considered above, which turns out to 
be $t_e$. Therefore, for the temporal duration of the neutrino production 
process we have 
$\sigma_t\simeq t_e\sim 7\times 10^{-16}\,{\rm s}$.  
Correspondingly, for the energy uncertainty of the produced neutrino and the 
length of its WP we find   
\be
\sigma_E\simeq t_e^{-1}
\simeq 1\,{\rm eV}
\,,\quad\quad
\sigma_x\simeq 
2\times 10^{-5}\,{\rm cm}\,. 
\label{eq:final}
\ee 

It has been shown in \cite{Giunti:2002xg,Beuthe:2001rc} 
that the temporal duration of the neutrino production process $\sigma_t$ 
is given by the time of overlap of the WPs of all the particles involved in 
neutrino production. Our approach, based on the consideration of mean free 
times of the involved particles, is in accord with this result.  
Our treatment merely implies that we take the lengths the WPs of these 
particles to be given by their mean free paths, 
which in fact determine their spatial localization. 
 
%%%%%%%%%%%%%%%%ffff1%%%%%%%%%%%%%%%%%%%%%%%%%%%%%%%%%%%%%%%
\begin{figure}
\centering
\includegraphics[width=0.7\linewidth]{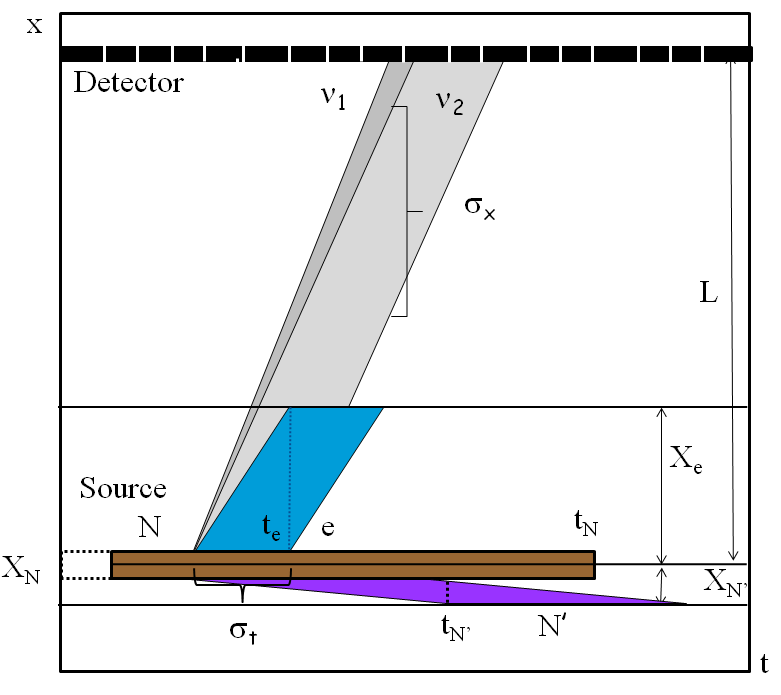}
\caption{
Schematic representation (space-time localization diagram) for 
neutrino production, propagation and detection in reactor experiments. Wave 
packets of the decaying nucleus $N$, daughter nucleus 
$N'$, electron and mass-eigenstate components of the emitted  
neutrino are represented by brown, violet, 
blue and gray bands, respectively.  
The slopes of the bands are determined by the group velocities of the particles.  
Black rectangles at $x = L$ show localization of the neutrino detection process. 
For simplicity, the WPs of only two neutrino mass eigenstates are shown.}
\label{fig:wp}
\end{figure}
%%%%%%%%%%%%%%%%%%%%%%%%%%%%%%%%%%%%%%%%%%%%%%%%%%%%%%%%%%%%%%%%%%%%%

This can be illustrated by space-time diagrams showing neutrino production, 
propagation and detection processes as well as the propagation and 
interactions of the accompanying particles in the WP approach 
\cite{Akhmedov:2012uu}. In Fig.~\ref{fig:wp} the propagation of the WPs of 
the particles is schematically represented by bands in the $(t,x)$ plane. The 
vertical sections of the bands give the lengths of the WPs, whereas their 
slopes are determined by their group velocities. The brown rectangle 
corresponds to the WP of the parent nucleus ($v_N\simeq 0$).  Due to their 
different group velocities, the WPs of the neutrino mass eigenstates ($\nu_1$ 
and $\nu_2$) diverge with time and traveled distance, and their overlap 
decreases. The horizontal sections of the bands corresponding to 
the produced $N'$ and electron are given by their mean free 
times, $t_{N'}$ and $t_e$, respectively; their intersection 
with the band 
representing the parent nucleus $N$ gives the space-time localization of 
the production process and, in particular, determines its duration $\sigma_t$. 
Thus, $\sigma_t$ is given by the overlap time of the WPs of all the particles 
participating in neutrino production and is dominated by the WP of the 
particle with the shortest mean free time.

%%%%%%%%%%%%%%%%%%%%%%%%%%%%%%%%%%%%%%%%%%%%%%%%%%%%%%%%%%%%%%%%%%%%%%%
\subsection{\label{sec:source2}WP lengths and neutrino spectra in 
radioactive source experiments 
}
%%%%%%%%%%%%%%%%%%%%%%%%%%%%%%%%%%%%%%%%%%%%%%%%%%%%%%%%%%%%%%%%%%%%%%%

Let us estimate the WP lengths $\sigma_x$ and the spectra $S(\bar{E})$ 
of neutrinos in source experiments, taking experiments with radioactive 
$^{51}$Cr as an example. Experiments with chromium source were carried out by 
GALLEX, SAGE and BEST collaborations. 

As was mentioned in section~\ref{sec:source1}, the shape and the width of the 
experimentally observed neutrino line are formed by two types of broadening 
effects -- homogeneous and inhomogeneous broadening. Homogeneous broadening 
forms the 
line of each individual emitted neutrino of a given mean energy 
$\bar{E}$; it includes such effects as natural linewidth, related to the finite 
lifetime of the parent atom, and collisional broadening due to the interaction 
of both the parent atom and of the particles accompanying neutrino production 
with the surrounding medium. These effects 
determine the width $\sigma_E$ of the neutrino WP in energy representation  
$f(E,\bar{E})$ and the coordinate-space neutrino WP length $\sigma_x\simeq 
v_g/\sigma_E$. It is the homogeneous line broadening that may lead to  
decoherence by WP separation in neutrino source experiments.  
 
Inhomogeneous broadening is related to each individual neutrino emitter being 
in somewhat different conditions, like slightly different energies of atoms in 
a crystal due to crystal defects or velocity spread due to the thermal 
motion of the source atoms. This leads to emission of neutrinos with slightly 
different mean energies $\bar{E}$, i.e.\ it forms their spectrum 
$S(\bar{E})$ discussed in section~\ref{sec:source1}. The spread of 
$\bar{E}$ due to inhomogeneous broadening has nothing to do with decoherence 
by WP separation; however, as discussed above, averaging over the mean neutrino 
energies may also lead to observable damping of neutrino oscillations and 
thereby mimic quantum WP decoherence effects. 
  
$^{51}$Cr decays via the electron capture process 
\be
^{51}{\rm C}r + e^- \rightarrow{}^{51}{\rm V} + \nu_e 
\label{eq:CrDecay}
\ee
with emission of four quasi-monoenergetic neutrino lines, grouped pairwise 
around $0.75$\,MeV (90\%) and 0.43\,MeV (10\%), with the half-life time 
$27.7$\,d. For definiteness, we will consider emission of neutrinos with 
energy 0.75\,MeV. For the electron capture process (\ref{eq:CrDecay}) the 
temporal duration $\sigma_t$ of the neutrino production process will be given 
by the smaller between the mean free times of the parent $^{51}$Cr and the 
produced $^{51}$V atoms.  

We shall first consider homogeneous broadening effects. 
Chromium is a transition metal with $bcc$ (body-centered cubic) crystalline 
structure and lattice constant $2.91\times 10^{-8}$\,cm.
As the localization length of the chromium atoms $X_{\rm Cr}$, one can take 
the rms deviation of their positions from the equilibrium positions in the 
crystal due to thermal vibrations.  
Assuming the temperature of the chromium source 
$T\sim 600$\,K,%
\footnote{Note that the temperature of the source decreases with 
time during the experiment.}, 
we find $X_{\rm Cr}\simeq 8.2\times 10^{-9}$\,cm~\cite{trampenau}. 
For rms velocity of thermal vibrations of chromium atoms 
in the crystal lattice we find 
$v_{\rm Cr}\simeq 1.7\times 10^{-6}c$. This gives the mean free time of 
chromium atoms $t_{\rm Cr}=X_{\rm Cr}/v_{\rm Cr}\simeq 
1.6\times 10^{-13}$\,s.

Consider now the mean free time of vanadium atoms. 
 {}From the kinematics of the decay it follows that the vanadium nuclei and 
the neutrinos produced in the decay process have equal momenta,  
$p_{\rm V}= p_\nu= 0.75$\,MeV. Therefore, the recoil energy and the velocity of 
the vanadium nucleus are $E_V\simeq 5.9$\,eV and $v_V \simeq 1.6 
\cdot 10^{-5}\,c$, 
respectively. Because the recoil velocity of vanadium is small 
compared with the velocity of atomic electrons, vanadium 
emerges from the decay process in the form of neutral or weakly ionized atoms. 
The mean free path of the vanadium atoms in chromium can therefore be found as  
$X_{\rm V}=(\sigma_{AA}n_{\rm Cr})^{-1}$, where $n_{\rm Cr}=8.6\times 
10^{22}$\,cm$^{-3}$ is the number density of chromium atoms and $\sigma_{AA}$ 
is the cross section of V--Cr atomic scattering. 
The latter can be estimated as the geometrical cross section 
$\pi(r_{\rm vdW,Cr}+r_{\rm vdW,V})^2\simeq 5.15\times 10^{-15}\,{\rm cm}^2$,  
where we have used the numerical values of the van der Waals radii of chromium 
and vanadium $r_{\rm vdW,Cr}=2.00\times 10^{-8}$\,cm and $r_{\rm vdW,V}=
2.05\times 10^{-8}$\,cm. For the mean free path and mean free time of the 
vanadium atoms we then find $X_{\rm V}=2.26\times 10^{-9}$\,cm, 
$t_{\rm V}=X_{\rm V}/v_{\rm V} \simeq 4.7\times 10^{-15}$\,s. 

Because the mean free time of the vanadium atoms is much shorter than that of 
the chromium ones, the temporal duration of the neutrino production process 
is determined by the former: $\sigma_t\simeq t_{\rm V}\simeq 4.7\times 
10^{-15}$\,s. Correspondingly, 
for the energy width and the length of the neutrino WP we obtain 
\be
\sigma_E\simeq t_{\rm V}^{-1}\simeq 
0.14\,{\rm eV}\,,\qquad\quad 
\sigma_x\simeq 
X_{\rm V}\frac{v_g}{v_{\rm V}}\simeq 
1.4\times 10^{-4}\,{\rm cm}\,. 
\label{eq:finalCr}
\ee 
Thus, the length of the neutrino WP is given by the mean free path of the 
vanadium atoms magnified by the factor $v_g/v_{\rm V}\simeq 6.3\times 10^{4}$. 
The contribution of the natural linewidth of $^{51}$Cr to $\sigma_E$ 
($\Gamma_{\rm Cr}\simeq 3.96\times 10^{-22}$\,eV) is completely negligible. 

Consider now the distribution of the mean energies of the neutrino WPs,  
$S(\bar{E})$. As was mentioned above, there are several incoherent broadening 
effects that contribute to it, such as e.g.\ crystal defects and impurities; 
the largest contribution comes from Doppler broadening related to the fact 
that the source atoms are not at rest but experience thermal vibrations with 
the rms velocity $v_{\rm Cr}$. Doppler broadening leads to the Gaussian shape 
of the line, as given in eq.~(\ref{eq:Gauss4}). In the case under 
consideration the corresponding width is 
\be
\delta_{El}=
\frac{v_{\rm Cr}}{c}E_0\simeq 1.3\,{\rm eV}\,, 
\label{eq:Dopp}
\ee
where we have used $E_0=0.75$\,MeV for the central energy of the line. 
Thus, the width of the energy spectrum of the $^{51}$Cr line $\delta_{El}$ 
exceeds the energy uncertainty $\sigma_E$ of the neutrino WPs 
by an order of magnitude.

Effect of localization of the accompanying particles ($N'$ from $e-$capture, 
$e$ and $N'$ from $\beta-$decay) on the neutrino WP length can be estimated 
in different way: Instead of the mean free paths of the accompanying 
particles one can consider the temporal durations of processes of their 
interactions with particles of medium $\sigma_t^i$ ($i = e, N'$). Velocities 
of the accompanying particles $v_i$ and $\sigma_t^i$ allow to construct the 
corresponding WP bands. Then the neutrino WPs will be 
determined by intersection of these bands and the band of original nuclei 
$N$. For $\sigma_t^i \ll t_N$, the size of neutrino WP will be substantially 
reduced in comparison to the case when interactions of the accompanying 
particles are neglected.

In turn, $\sigma_t^i$ are determined by WPs of atoms of medium and WPs of 
products of secondary particle interactions. The problem is that one should 
consider the chain of interactions which probably ends up by thermalization 
of the products.

In the case of $e-$capture the WP size of produced Vanadium is determined by 
the duration of the process $^{51}V + ^{51}Cr\rightarrow ^{51}V'+ 
^{51}Cr'$, $\sigma_{Cr}$. To find $\sigma_{Cr}$, apart from localization 
of $^{51}Cr$, one should know the WPs of $^{51}V'$ and $^{51}Cr'$, etc. The 
estimation show that the chain of secondary interactions can reduce 
$\sigma_{Cr}$ by factor $10^{-2} - 10^{-1}$ in comparison to $t_{Cr}$. This 
gives the length of neutrino WP, $\sigma_x$, of the same order as in 
(3.18). 

The approach, based on the consideration of mean free times, is in fact a 
shortcut allowing us to avoid consideration of  the ladder of 
interactions of daughter particles. This approach uses mean free times  
for localization of particles universally: both for the parent particle 
$N$ and the accompanying particles.

A note on the detection processes is in order. As was pointed out 
above (see footnote~\ref{ftnt:det}), our estimates show that for reactor and 
source experiments energy uncertainties inherent to neutrino detection are  
much larger than those inherent to neutrino production, $\sigma_E^D \gg 
\sigma_E^S$, so that to a very good accuracy $\sigma_E=\sigma_E^S$.
The condition $\sigma_E^D \gg \sigma_E^S$ allows one to replace $f_D(E)$ 
in eq.~(\ref{eq:f1}) by a constant; this means that in the cases under 
consideration the detection processes do not affect neutrino coherence. For this 
reason we did not consider them in detail.

\section{\label{sec:discuss}Discussion}

We have shown that the effects of decoherence by WP separation can always 
be incorporated into a modification of the detector resolution function or, 
for source experiments, of the shape and width of the neutrino line; therefore, 
these two sources of the oscillation damping are equivalent from the 
observational point of view. 
Note that in the case of Gaussian averaging, the observational equivalence of 
WP decoherence and averaging over $L/E$ was previously shown in 
\cite{Ohlsson:2000mj}. 
 
We have found that the effective detector energy resolution functions (or 
effective widths of the neutrino line) are always dominated by the larger 
between the inherent QM uncertainty of neutrino energy $\sigma_E$ and 
the detector resolution $\delta_E$ (or the linewidth $\delta_{El}$). 
For Gaussian WPs and resolution functions, the effective resolution is 
characterized by
\be
\delta_{E{\rm eff}}=\sqrt{\delta_E^2+\sigma_E^2}\,,
\label{eq:eff} 
\ee
and similarly for the effective neutrino linewidth $\delta_{El{\rm eff}}$ 
for the source experiments. It should be stressed that in the latter case the 
width of the neutrino spectrum $S(\bar{E})$ is dominated by Doppler broadening, 
which does lead to $S(\bar{E})$ of Gaussian form.  

Our results show that for reactor neutrinos the effects of WP separation may 
only be experimentally probed if $\sigma_E > \delta_E$ (for 
radioactive source experiments the corresponding condition is $\sigma_E > 
\delta_{El}$). This has to be complemented by a condition on the baseline of the 
experiment $L$, which will be discussed below.% 
\footnote{
From eq.~(2.20) one could conclude that WP separation effects may be 
observable even if the condition $\sigma_E \gtrsim \delta_E$ is not met,  
provided that the energy resolution of the detector is known with very high 
accuracy, so that its error $\Delta(\delta_E)$ is smaller than 
$\sigma_E$. However, as follows from~(2.20), this would require 
unrealistically high accuracy of the detector resolution function, 
$\Delta(\delta_E)\sim \sigma_E^2/\delta_E^2$; more importantly, as we shall see, 
even perfectly known energy resolution of the detector would not allow one to 
observe WP separation effects because of the constraints on the baseline $L$.}

For reactor experiments, our estimates gave for the lengths of the neutrino WPs 
and the corresponding QM neutrino energy uncertainties 
\be
\sigma_x\simeq 2\times 10^{-5}\,{\rm cm}\,,\qquad
\sigma_E\simeq 1\,{\rm eV}\,.
\label{eq:fin1}
\ee
The latter value has to be compared with detector energy resolution in reactor 
experiments. Currently, it is in the sub-MeV region; the forthcoming JUNO 
experiment aims at a very high energy resolution of about 3\%, which 
corresponds to $\delta_E\sim 100$\,keV. Eq.~(\ref{eq:fin1}) then gives 
$\sigma_E/\delta_E\sim 10^{-5}$, which means that effects of oscillation 
damping by WP separation cannot be seen in reactor neutrino experiments. 
This also means that these WP-related quantum decoherence effects   
cannot hinder the determination of the neutrino mass ordering, which is one 
of the major goals of JUNO. 

Our estimate of the WP lengths of reactor neutrinos in eq.~(\ref{eq:fin1}) can 
also be compared with the lower bound (\ref{eq:combined}) 
obtained from the combined analysis of the existing reactor 
neutrino data in \cite{deGouvea:2020hfl,deGouvea:2021uvg}; 
our result exceeds this lower limit by six orders of magnitude. 

For the WP lengths and intrinsic energy uncertainties of 
neutrinos in chromium radioactive source experiments 
we found 
\be
\sigma_x\simeq 1.4\times 10^{-4}\,{\rm cm}\,,\qquad
\sigma_E\simeq 0.14\,{\rm eV}\,.
\label{eq:fin2}
\ee
These values differ from the corresponding values for reactor neutrinos by 
roughly one order of magnitude. 
The value of $\sigma_E$ in eq.~(\ref{eq:fin2})  
is about a factor of ten smaller than the energy width of the detected 
neutrino line, $\delta_{El}\simeq 1.3$\,eV. We see that the disparity between 
the QM neutrino energy uncertainty and the energy resolution of the experiment 
is in this case much weaker than it is for reactor experiments.

As was pointed out in the Introduction, in ref.~\cite{Arguelles:2022bvt} an 
attempt had been made to reconcile the results of short-baseline reactor 
experiments and the radioactive source experiment BEST basing on the WP nature 
of neutrinos. The value of the neutrino WP length $\sigma_x$ was chosen 
to be equal to the lower bound (\ref{eq:combined}) found in 
\cite{deGouvea:2020hfl,deGouvea:2021uvg}. However, according to our estimates, 
this lower bound is orders of magnitude below the actual values of $\sigma_x$ 
for both reactor and chromium-source neutrinos.  
Therefore, WP separation effects cannot reconcile the results of the reactor 
and BEST experiments. 

Can one still contemplate a terrestrial experiment that would be sensitive to 
the size of the neutrino WP? As $\sigma_E/\delta_{El}$ is relatively large 
for $^{51}$Cr source experiments, one could think about 
source experiments with smaller $\delta_{El}$, e.g.\ 
look for radioactive sources with smaller 
neutrino energy $E_0$. This would reduce the Doppler broadening of the 
neutrino line, which is proportional to $E_0$ (see eq.~(\ref{eq:Dopp})). 
However, decreasing $E_0$ would also mean that the recoil momenta of the 
daughter nuclei  produced in the electron capture process would decrease. 
As follows from the discussion in section~\ref{sec:source2}, this would 
also decrease $\sigma_E$; as a result, the ratio $\sigma_E/\delta_{El}$ will 
be unaffected.  

A more practical option would probably be to cool down the source.  This would
suppress Doppler broadening effects without decreasing $\sigma_E$. 
In any case, for radioactive source experiments the values of $\sigma_E$ and 
$\delta_{El}$ are not very different, does that mean that 
one could probe decoherence by WP separation in such experiments? 

The answer seems to be negative. The point is that although the requirement  
$\sigma_E\gtrsim \delta_E $ (or $\sigma_E\gtrsim \delta_{El})$ 
is a necessary condition for observability of WP separation effects, it is 
not sufficient. Irrespectively of whether $\sigma_E$ is smaller or larger than 
$\delta_E$ (or $\delta_{El})$,  
for the WP separation effects to develop neutrinos should travel sufficiently 
large distances, comparable to or larger than the coherence length. 
For reactor neutrino experiments, from eq.~(\ref{eq:fin1}) and the definition 
of the coherence length in eq.~(\ref{eq:Lcoh}) we find 
\be
L_{{\rm coh},21}\simeq 
4.8\times 10^{7}\,{\rm km}\,,\qquad 
L_{{\rm coh},31}\simeq 
1.4\times 10^{6}\,{\rm km}\,,\qquad 
L_{{\rm coh},41}\simeq 
3600\,{\rm km}\,.
\label{eq:conL}
\ee
Here $L_{{\rm coh},21}$ and $L_{{\rm coh},31}$ correspond to neutrino mass 
square differences $\Delta m_{12}^2\simeq7.5\times 10^{-5}\,{\rm eV}^2$ and 
$\Delta m_{31}^2\simeq 2.5\times 10^{-3}\,{\rm eV}^2$, respectively, and  
$L_{{\rm coh},41}$ corresponds to much discussed hypothetical 
active-sterile neutrino oscillations with 
$\Delta m_{41}^2\simeq 1\,{\rm eV}^2$. 
Neutrino energy $E=3$\,MeV was assumed. 
For the chromium source experiment ($E=0.75$\,MeV), the value of 
$\sigma_x$ from eq.~(\ref{eq:fin2}) yields 
\be
L_{{\rm coh},21}\simeq 
2.1\times 10^{7}\,{\rm km}\,,\qquad 
L_{{\rm coh},31}\simeq 
6.3\times 10^{5}\,{\rm km}\,,\qquad 
L_{{\rm coh},41}\simeq 
1600\,{\rm km}\,.
\label{eq:conL2}
\ee
Obviously, no reactor or neutrino source experiments with such baselines 
are possible. 

The coherence lengths are inversely proportional to $\Delta m_{ik}^2$, and 
one could therefore expect that it is easier to probe the effects of 
decoherence by WP separation in experiments that are sensitive to larger mass 
square differences, such as active-sterile neutrino oscillations experiments 
\cite{Arguelles:2022bvt}. This is, 
however, misleading. 
The point is that experiments are usually devised such 
that the experimental baseline is of the order of the expected neutrino 
oscillation length. As the latter is also inversely proportional to 
$\Delta m_{ik}^2$, the ratio 
\be
\frac
{L_{{\rm coh},ik}}{l_{{\rm osc},ik}}=
\frac{\sigma_x E}{2\pi}
\label{eq:ratio}
\ee
is independent of $\Delta m_{ik}^2$. Note that this quantity is 
Lorentz-invariant, as so is $\sigma_x E$ \cite{Farzan:2008eg}. 
For reactor experiments we find $L_{\rm coh}/l_{\rm osc}\sim 5\times 10^5$, 
that is, decoherence by WP separation would have started to be seen only 
after neutrinos have propagated half a million oscillation lengths. 
Similar estimate holds for neutrino source experiments. 
This is by far much larger than any reasonable baseline in terrestrial neutrino 
experiments. Obviously, even if experiments with such huge $L$ were possible, 
effects of averaging due to finite energy resolution of the detectors 
would reveal themselves much before. From eq.~(\ref{eq:ratio}) it follows that 
the WP separation effects should become more pronounced with decreasing $E$;
it is however not easy to detect neutrinos with energies much below the 
MeV range.  

We conclude that it is not possible to observe effects of quantum decoherence 
by WP separation in terrestrial neutrino experiments, at least for the class 
of experiments we considered (i.e.\ reactor and radioactive neutrino source 
experiments). If a damping of the oscillations which exceeds the expected \
damping related to the usual averaging due to the (accurately known) 
finite energy resolution of the detector, or due to the finite neutrino 
linewidth in source experiments, is nevertheless observed, 
this will signify some new physics and not decoherence by WP separation.

\bigskip
{\it Acknowledgement.}
We thank D.~Gorbunov for useful correspondence. 

%%%%%%%%%%%%%%%%%%%%%%%%%%%%%%%%%%%%%%%%%%%%%%%%%%%%%%%%%%%%%%%%%

\end{document}